# How internal waves could lead to wreck American and Indonesian submarines?

Yury Stepanyants

School of Sciences, University of Southern Queensland, Toowoomba, Australia;

Department of Applied Mathematics, NSTU n.a. R.E. Alekseev, Nizhny Novgorod, Russa

The disaster that occurred in April 2021 with the Indonesian submarine in the Bali Sea again reminded to experts of the possible role of internal waves in such tragedies [1]. The hypothesis that internal waves could have caused the death of the American nuclear submarine "Thresher" in April 1963 has been discussed in scientific and popular literature many times. In this article, the author expresses his point of view on these events, and also provides additional data on the death of another American nuclear submarine "Scorpion" in May 1968. In order to understand what could have happened to the submarines, it is necessary at least briefly explain the physics of internal waves that exist almost everywhere inside the oceans, seas and other large basins.

**Surface wave physics**

Waves on the surface of water bodies are well known to everyone; their existence is associated with the restoring action of Earth's gravity. Therefore, such waves are called *gravity waves* (they, however, should not be confused with gravitational waves in the theory of relativity). If a part of the liquid surface for some reason is taken out of the equilibrium position, then the force of gravity tends to return the disturbed area to the equilibrium position. However, the liquid particles, returning to equilibrium, gain speed and, by inertia, slip past the equilibrium position, falling below the free surface, as shown in Fig. 1. Then a buoyant force (Archimedes' force) acts on them, forcing the particles of the liquid to move upward and again slip through the equilibrium position due to inertia and process repeats. At such motion, nearby particles are involved in the oscillatory motion capturing more and more distant regions from the initial disturbance and wave propagation from the place of the initial disturbance is initiated. Under certain conditions, if the displacement of liquid particles is suitably matched and appropriate velocities are given to them, it is possible to obtain a directed wave motion and, in particular, a single wave travelling in a given direction as a localised disturbance that does not change its shape in the course of propagation (such *stationary moving* waves are called *solitons*). A soliton was first observed in the canal between Edinburgh and Glasgow by the Scottish engineer John Scott Russell in 1834, who left a colourful description of this phenomenon (see, for example, in [2–4]).

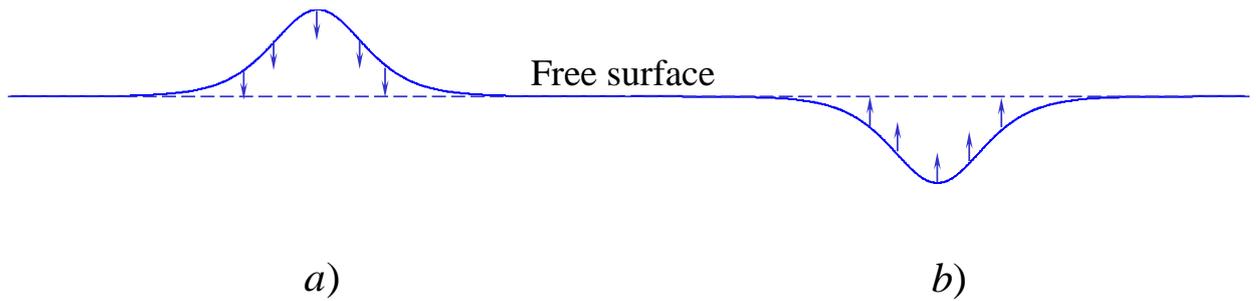

*Fig. 1. a) The elevation of the free surface of a liquid caused by an external disturbance. b) The lowering of the level of the free surface after liquid particles have passed their equilibrium position, shown by the dashed line. The arrows indicate the directions of the action of gravity forces on liquid particles.*

A team of scientists using modern means have demonstrated how a solitary wave in the same canal can be created, similar to that observed by Scott Russell (see Fig. 2). It should be noted that solitary surface waves can only be of positive polarity, i.e. in the form of an elevation of the liquid surface, as in Fig. 2.

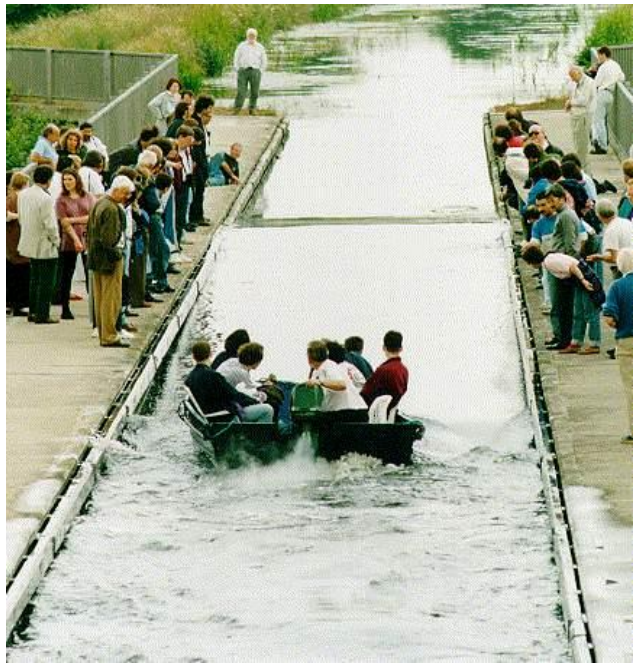

*Fig. 2. A solitary wave (soliton) created in a canal by a fast-moving motor boat (from free access on the Internet – see, for example, [4]).*

**Internal waves**

Wave motions can occur not only on the surface of a liquid, but also inside it at the interfaces of layers of different densities (*pycnoclines*) or even in the bulk when a fluid density smoothly varies with the depth. Such waves are called *internal waves*; they exist in a stably stratified fluid when the lighter layers are above the heavier ones. The physics of these waves is the same as for the surface waves described above – the displacement of a pycnocline from the equilibrium horizontal level produces forces that tend to return the particles of a heavier liquid to their levels below the pycnocline. If the lighter particles are below the interface between the layers, then the buoyancy Archimedes force acts on them, forcing them to move upward. However, the forces of gravity and buoyancy are significantly weakened inside the fluid, since instead of the usual gravity $g$, the reduced force $\Delta\rho g/\rho$ acts on them, where $\rho$ is the average density of the liquid layers, and $\Delta\rho$ is the difference in the densities of the lower and upper layers, which is usually small, $\Delta\rho/\rho = 10^{-3} - 10^{-4}$. Therefore, the frequencies of oscillations and the speed of propagation of internal waves are approximately 30–100 times less than those of surface waves. The density of water in the ocean mainly depends on the temperature and salinity. The upper layers are usually warmer (in some places the temperature can attain 20°C and even higher) and less salty, while at depths of more than 300–500 m the temperature is almost the same everywhere and is equal to 4° C.

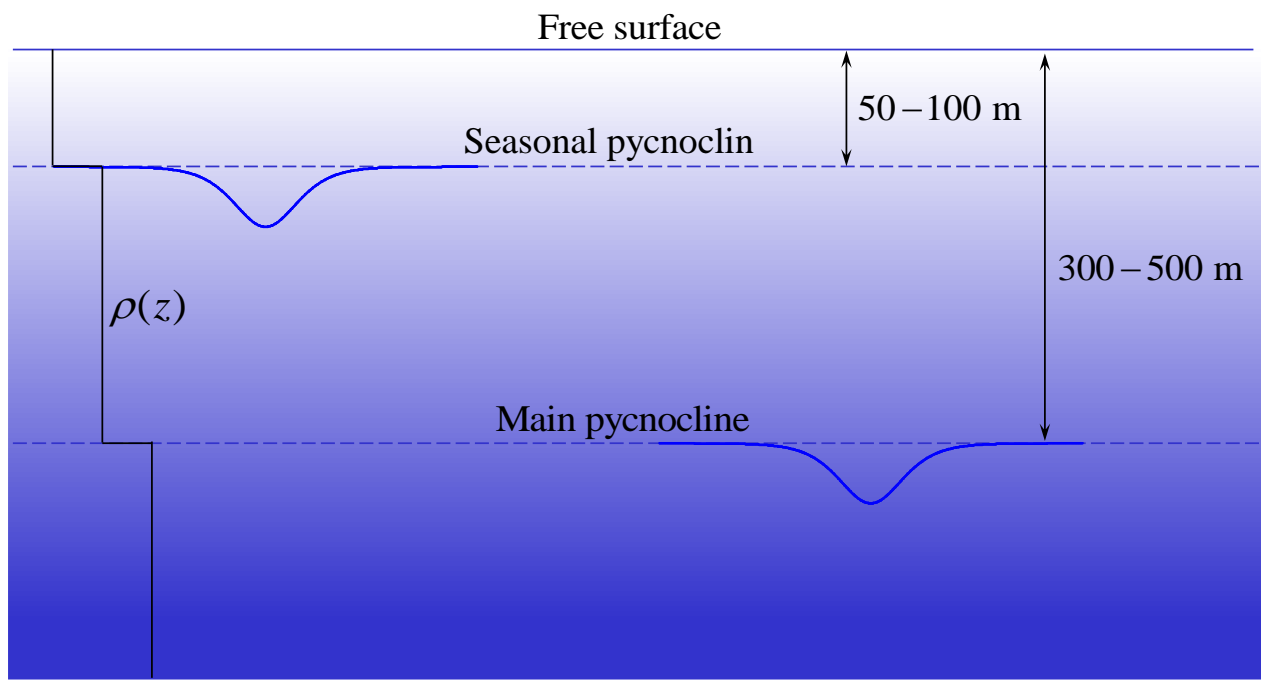

*Fig. 3. A simplified three-layer model of stratification of the World ocean with the abrupt change of density $\rho(z)$. In fact, instead of sharp density jumps in the ocean, a noticeably pronounced but smoothed density transition from layer to layer occurred as indicated by the continuous variation in shading.*

A simplified model of the water density variation with the depth in the ocean can be represented as shown in Fig. 3. The upper freshened layer, the thickness of which ranges from 50 m to 100 m, depending on the season, is formed due to the influx of solar heat, precipitation, and wind mixing. It is followed by a fairly well-pronounced jump in the density, called the *seasonal pycnocline*. Then the next layer of almost constant density extends, at the lower boundary of which a rapid change in density occurs again, after which the density of the liquid remains practically constant down to the bottom. The boundary separating the lowest ocean layer from the intermediate layer is called the *main oceanic pycnocline*. It exists in the ocean all the time, regardless of weather or seasonal changes on the surface.

Internal waves of rather large amplitude, up to 150 m, can exist at both interfaces of the layers. Moreover, such waves hardly disturb the free surface; it is almost impossible to see them if the ocean is completely calm and there are no wind waves. Figure 4 illustrates the excitation of internal waves in a laboratory flume by a slowly moving boat model. The tank is filled with a liquid consisting of three layers of different densities; the layers are tinted by dyes. It is clearly seen that the moving model excites internal waves at the interface between the layers, leaving the surface of the liquid unperturbed.

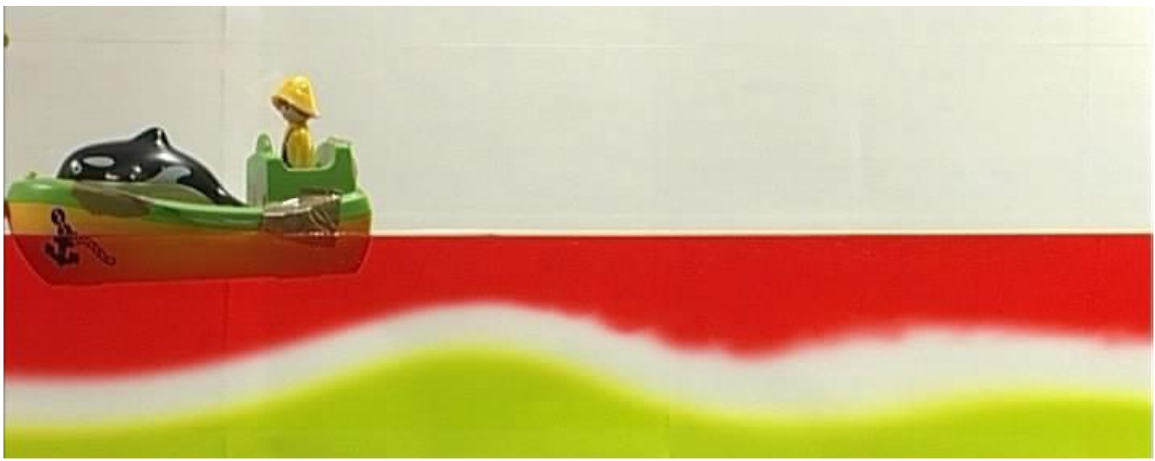

*Fig. 4. Excitation of internal waves in a laboratory flume by a slowly moving boat model [5]. The top layer (red) contains the light liquid, the middle layer (white) contains the heavier liquid, and the bottom layer (light green) contains the heaviest liquid.*

More details about the nature of internal waves can be found in the article by J. Apel [6] (in the section Background and Theory) (see also [7]). A specific feature of internal solitary waves (ISWs) in the deep ocean is that, unlike surface waves, they usually have negative polarity, i.e. they are exhibited by a moving *depression* of the pycnocline, as shown schematically in Fig. 3.

In cases where a pycnocline is close to the bottom, which often happens in the shelf areas, the polarity of solitons is, however, positive, i.e. they are exhibited by humps of the pycnocline as in the case of surface waves. The general rule of thumb is that they always point towards the thicker liquid layer. For surface waves, the thick layer is the atmosphere; surface waves can be considered as a special case of internal waves at the water-air interface.

In the presence of surface wind waves, internal waves, creating flows in the near-surface layer of the liquid with velocities up to 20 cm/s (which is comparable to the velocities of wind ripple propagation), produce modulation of surface waves. On the steep section of the front of an internal wave moving in the deep ocean, there is an increase in surface ripples, whereas on the rear slope of the wave there is a partial suppression of waves (see Fig. 5).

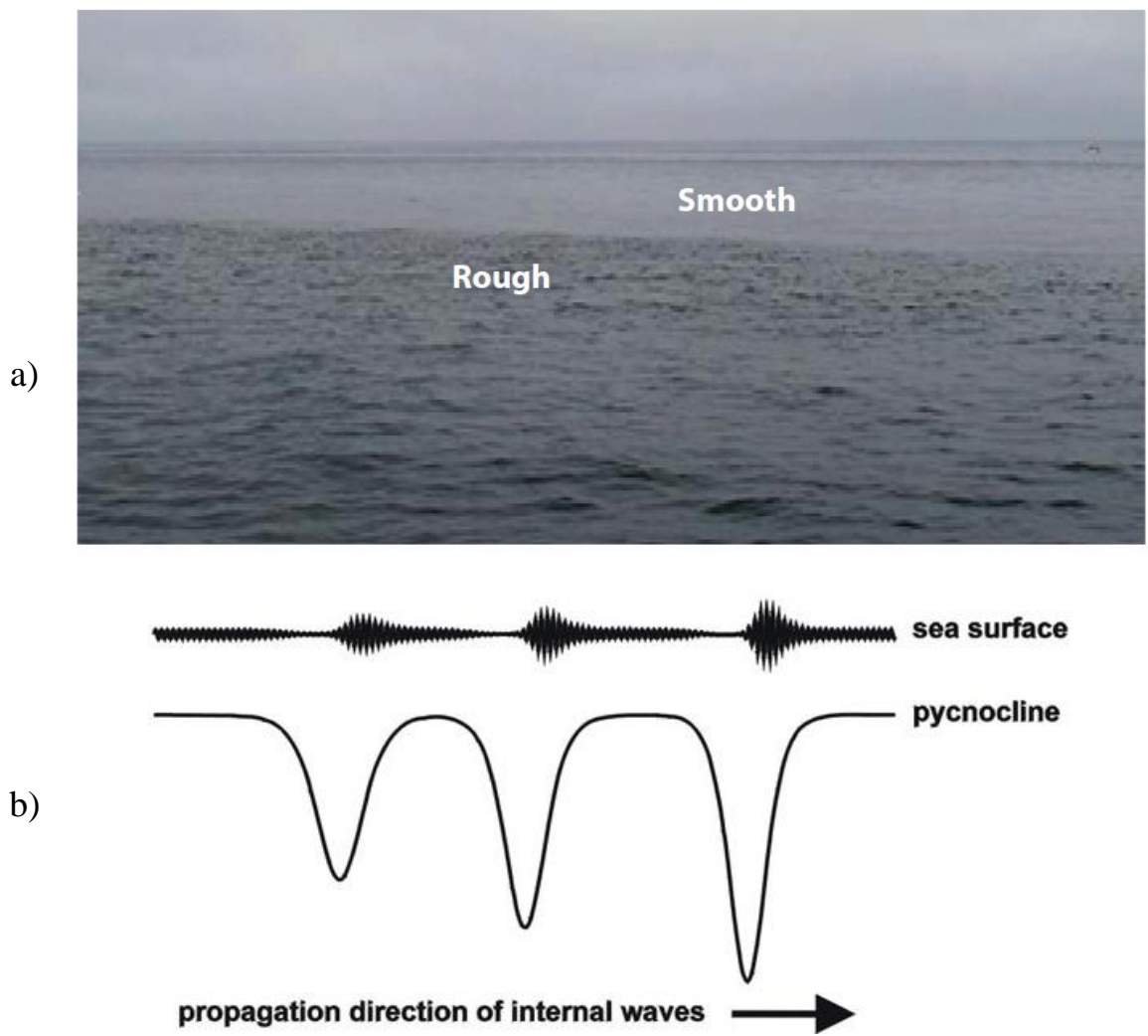

*Fig. 5. Photo of the ocean surface (a) with areas of increased roughness and suppressed waves above an internal wave slowly moving in the water [9]. Increased surface wave excitation occurs above the descending wave slope, and suppressed – above the rising wave slope, as shown in figure (b) from [7]. The arrow shows the direction of propagation of internal waves.*

As a result, stripes extending for many kilometres appear on the surface of the ocean, moving along with an internal wave propagating in the bulk of water. These stripes are clearly visible to the naked eye and are even observed in photographs obtained from satellites (see, for example, Fig. 6). The Internet is replete with photographs of this kind, see, for example, atlases [6–8].

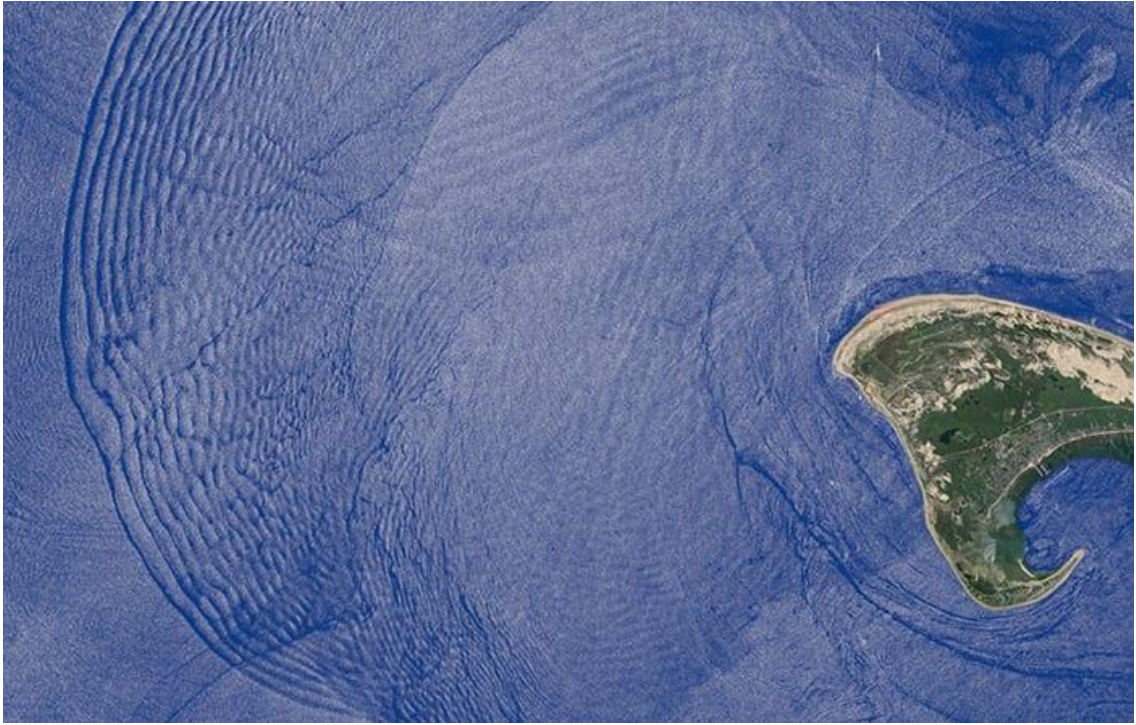

*Fig. 6. Satellite photograph of alternating light and dark stripes formed on the ocean surface by internal waves [10]. Waves travel from the Atlantic Ocean to the eastern coast of United States through the Gulf of Maine skirting Cape Cod.*

Currently there is a well-developed theory of internal solitons well-confirmed by numerous observations, laboratory experiments, and numerical calculations; details can be found in reviews [11–15]. Early models describing ISWs were based on the well-known Korteweg – de Vries (KdV) equation [2, 11–15], the soliton solutions of which were found in a fairly good agreement with the observed oceanic waves of moderate amplitudes (see Fig. 7). However, subsequent studies have shown that the ISW shapes can be very different from the bell-shaped KdV solitons shown in Fig. 5b) and 7. For their adequate description, either the generalized KdV equation, called the Gardner equation, or the complete sset of hydrodynamic equations should be used. In the former case, the Gardner equation can be solved analytically; as the result, the shapes of solitons were derived for different amplitudes (see Fig. 8a). In the second case when the complete set of hydrodynamic equations is used, shapes of solitons can be calculated numerically. The numerical solutions were obtained, and their good agreement with the analytical solution of the Gardner equation has been demonstrated.

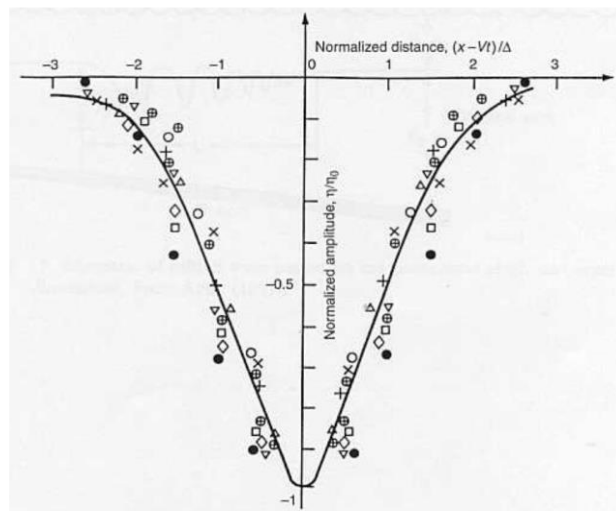

*Fig. 7. Comparison of the shape of the soliton solution of the KdV equation with the shapes of recorded ISWs in the Sea of Okhotsk [16]. The solid line is the analytical solution of the KdV equation, normalized to the maximum deviation from equilibrium; the symbols refer to the normalised values of various ISWs.*

An important result of this study is the conclusion that solitons of large amplitude in the deep ocean become similar to trenches with a flat bottom (see inset to Fig. 8a), whereas in shallow-water basis, for example, on a shelf with a near-bottom pycnocline, they are similar to elevations with a flat top. Experimental studies have shown that this type of ISWs is indeed observed in the oceans and seas. In particular, Fig. 8b) illustrates the recorded pycnocline perturbations on the shelf of the Sea of Japan [18]. The amplitude of the solitary wave #4 was 3 m whereas the depth of the shelf was 30 m, and the pycnocline was located at the depth of 25 m. As one can see, the shapes of pulse disturbances #1, #2, and #4 are close to the theoretically predicted curve 3 in Fig. 8a), whereas the shape of the smaller-amplitude pulse #3 is close to the shape of a KdV soliton (cf. with line 1 in Fig. 8a).

Parameters of ISWs depend on the geographic location of observation, ocean depth and the forces which generate the waves. One of the most important factors in the generation of internal waves (but not the only one) is the semidiurnal tide. In some areas of the ocean (for example, in the Andaman Sea, the Sulu Sea, the South China Sea, and in various locations in Indonesian waters), internal waves of large amplitude appear regularly – approximately every 12.5 hours, after high tides. Especially large disturbances arise when the lunar and solar tides coincide in time (being in a *syzygy*). Below is a Table from the review [14], in which the data on the main

ISW parameters (amplitudes, velocities, characteristic sizes and durations, as well as front lengths) were systematized at various ocean depths. As can be seen from this Table, in deep-water areas of the ocean with depths of more than 1,000 m, the amplitudes of solitons can attain 130 m, and according to some data, even 150 m. The characteristic width of solitons can attain 2,000 m, and the propagation velocity 2.5–3 m/s (9–11 km/h) with a front length of up to 350 km.

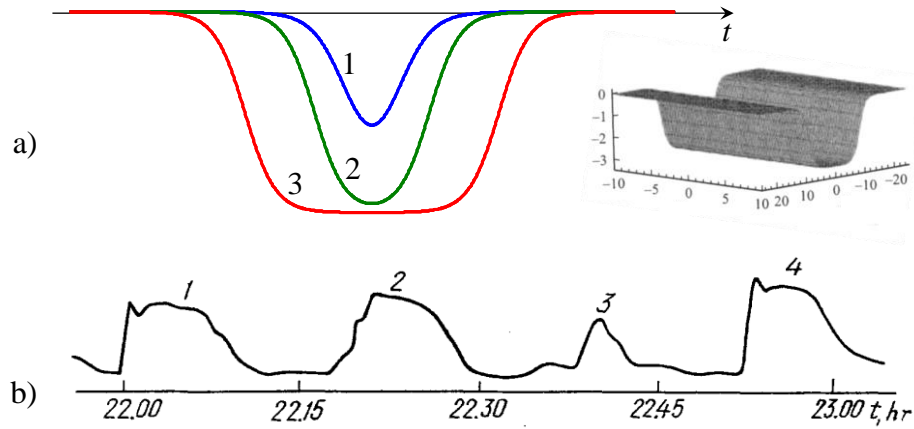

*Fig. 8. a) Theoretically predicted shapes of solitons in the deep ocean within the framework of the model Gardner equation at various amplitudes [17]. Curve 1 refers to small and moderate amplitude bell-shaped solitons; curve 2 – to large-amplitude solitons; curve 3 – to solitons of very large amplitude (scales along the horizontal and vertical axes are not shown, the figure is of a qualitative nature). b) Experimentally measured pycnocline perturbations on the shelf of the Sea of Japan [18] (the amplitude of the solitary wave #4 was ~3 m).*

|  | Amplitudes, m | Widths, m | Speeds, m/s | Other parameters |
|---|---|---|---|---|
| Shallow basins ($H \sim 4$—20 m) | 3—5 | 65—250 | 0.4—0.8 |  |
| Medium depth seas and lakes ($H \sim 20$—1000 m) | 5—20 | 200—400 | 0.5—1.5 | Total number of solitons in the train – 20 |
| Deep seas and ocean ($H \sim 1$—5 km) | 20—130 | 200—2000 | 1.0—2.5 | Length of wave fronts 100—350 km |

*Table 1. Parameters of ISWs in different regions of the oceans and other basins [14].*

The review [11] presented for the first time a map of ocean regions in which ISWs were observed. Subsequently, this map was significantly extended by C. Jackson and is now publicly available on the website [6] (see also [7, 8]). In Fig. 9 from the atlas [6] it can be seen that ISWs are observed almost everywhere in the oceans. The density of points on the map is only due to the convenience of conducting experimental work in certain areas of the World Ocean. The lack of data on the Pacific and Southern Oceans is due to the relative inaccessibility of the regions, the high cost of experimental work there, and lack of intensive navigation in them. Therefore, it is not known what surprises can still be expected in these vast and poorly studied areas of the World Ocean.

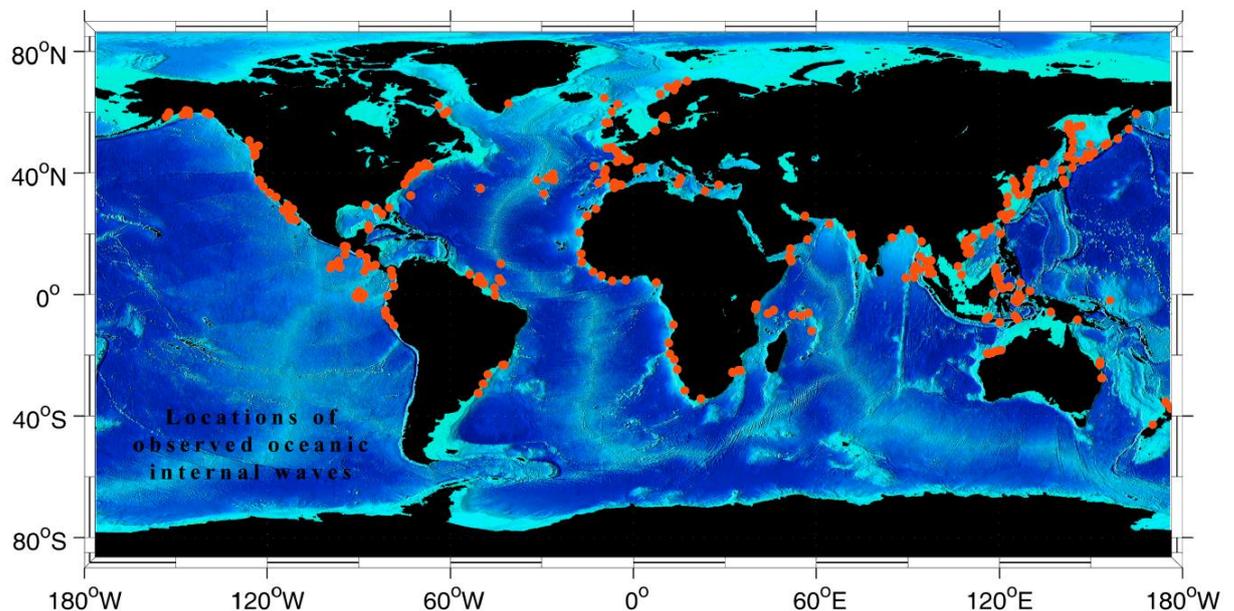

*Fig. 9. A map from the Jackson's atlas [6] depicting the sites of observation of large-amplitude ISWs.*

What danger can ISWs pose to submarines? Let's imagine that a submarine with a displacement of 4,300 tons (which roughly corresponds to the lost American submarine "Thresher") moves under the main pycnocline located at a depth of 300 m, where the density at a temperature of 4°C and salinity 35‰ (35 ppm) is $\rho_2 = 1{,}027.8$ kg/m$^3$. The submerged displacement is defined as the weight of the liquid displaced by the body, therefore, with the indicated displacement, the volume of the submarine is $V \approx 3{,}375$ m$^3$. The submarine is in a neutral, suspended state so that the force of gravity is compensated by the Archimedean buoyancy force. Suppose it suddenly turns out to be in a water of lower density $\rho_1 = 1{,}027.4$ kg/m$^3$ (these values correspond to the data on stratification at the site where the nuclear submarine "Thresher" sank [19]). Then, the Archimedean force decreases so that an additional gravity force $f = (\rho_2 - \rho_1) g \approx 3{,}9$ kg/(m$^2$s$^2$) acts on each cubic meter of the submarine, where $g = 9.81$ m/s$^2$ is the acceleration due to gravity. Therefore, an excess force of $F = fV = 1.3 \cdot 10^4$ kg·m/s$^2$ acts on the entire submarine, which corresponds to a load of 1.35 tons. With such a load, the submarine will begin to submerge and can quickly reach the maximum depth, at which is designed for its durable body, if you do not take emergency measures for its ascent. Here it is appropriate to give simple estimates for the submarine model in the form of an ellipsoid of revolution around its long axis. Let us assume that the submarine, moving horizontally, suddenly passes from the region under the pycnocline with the density $\rho_2$ to the core of ISW with a lower density $\rho_1$, being at the same horizon (see Fig. 10). Then, in the simplest case, its vertical coordinate $z$ as a function of time $t$ is determined by the differential equation:

$$(\rho_s + \rho_0 k) \ddot{z} = -(\rho_2 - \rho_1) g,$$

where $\rho_s$ is the average density of a submerged submarine that is equal to the average density of upper and lower layers $\rho_s = \rho_0 \equiv (\rho_2 + \rho_1)/2$, $k = 0.933$ is the coefficient of the added mass of an ellipsoid with the aspect ratio of 7 when it moves vertically perpendicular to its longitudinal axis, and the dots above $z$ stand for time differentiation (here, for estimates, the submarine is replaced by an ellipsoid of revolution around the longitudinal axis).

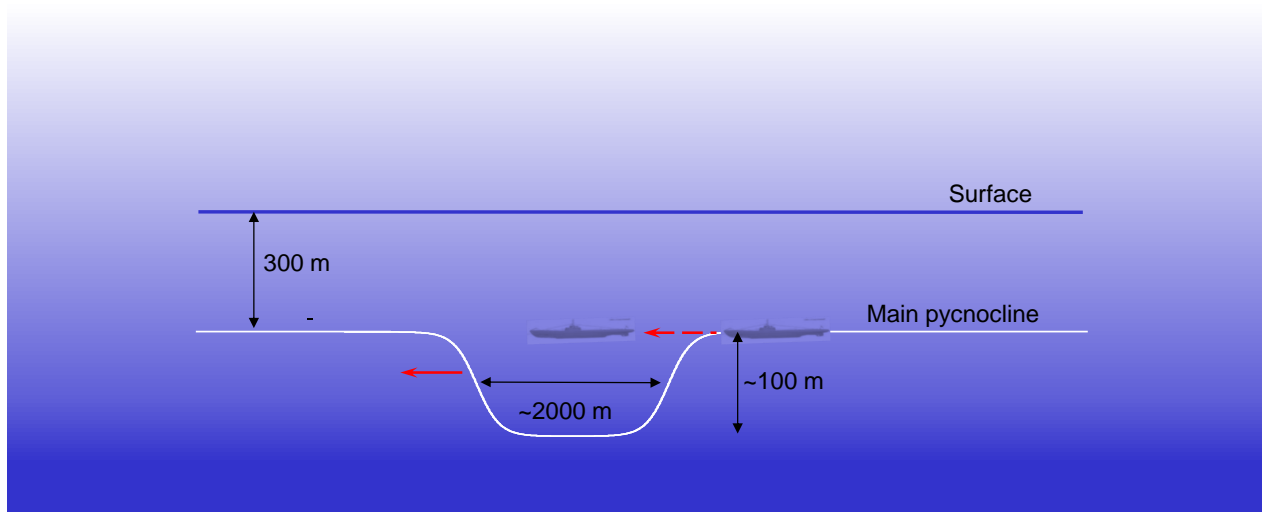

*Fig. 10. Schematic representation of a submarine overtaking a solitary wave over the main pycnocline. The drawing is not to scale.*

The solution to the above equation with the zero initial conditions is a simple function of time: $z(t) = a\,t^2/2$, where $a = -(\rho_2 - \rho_1)g/(\rho_s + k\rho_0)$ is the acceleration of a submarine submersion due to the lack of buoyancy. If we substitute here the above values of the parameters, we get $a \approx -0.002$ m/s$^2$. Suppose that the submarine moves at a speed of 6 knots ($\approx 3$ m/s) catching up an ISW of 100 m amplitude. Such a wave can travel at a speed of 2–2.5 m/s (see the Table above) having a characteristic width of about 1,000 m. To overtake the ISW with a speed difference of 1.5 m/s, the submarine will need 667 seconds (11 minutes). During this time, with the indicated acceleration, the submarine can theoretically submerge to an additional 445 m if urgent actions are not taken to keep it at a given horizon. In fact, a submerged submarine sinking additionally $H = 100$ m from the pycnocline horizon of 300 m may already turn out to be catastrophic. The time required for such a dive at the given parameters is $t_s = (2H/a)^{1/2}$, which is 316 sec (5.3 min).

It is appropriate to recall here the air pockets which sometimes huge modern airliners fall into. As a result, in some cases the plane falls hundreds of meters down (sometimes even up to several kilometres). However, at the same time, the airplane body can easily withstand changes in atmospheric pressure by only a few percent. This is not the case with a submarine the external pressure on the hull of which greatly increases with the depth – one additional atmosphere for every 10 m. To keep a submarine at a given horizon, it is necessary to urgently take measures to compensate for the excess force; for this it is necessary to purge the main ballast tanks or control the rudders. The submarine under water is in a dynamic balance with the environment, and its maintenance in a state of neutral buoyancy at a given depth requires constant corrections. They are produced by horizontal rudders or by pumping/receiving seawater as a ballast; this is called a submarine sign or depth stabilization. As follows from the above, a sudden loss of buoyancy can occur if a submarine enters a depression type ISW which has a negative polarity and contains water of reduced density. Such an event becomes extremely dangerous if the submarine is already at a depth close to the limiting one or even at a working depth which is usually of 70% – 90% of the limiting one.

**Analysis of possible reasons for the mysterious wrecks with three submarines**

Let us analyse three well-known disasters that happened to submarines the causes of which have not yet been established. The first of these major accidents occurred with the US nuclear attack submarine "Thresher" in April 1963. The second happened with another US nuclear submarine "Scorpion" in May 1968. And finally, the recent disaster in April 2021 with the diesel submarine of the Indonesian Navy "Nanggala-402". Below are the discussions of the accidents that have occurred in chronological order and attempts to shed a light on the wrecks.

*The wreck of the nuclear submarine "Thresher", April 1963. The first disaster in the history of the world nuclear submarine fleet.*

Reportedly, the "Thresher" nuclear submarine was repaired and tested in the Gulf of Maine off the east coast of the United States. The nuclear submarine "Thresher" (a thresher is a certain type of shark) was launched from the Portsmouth shipyard in 1958. At that time, "Thresher" was the most modern attack submarine of the US Navy distinguished by high maneuverability with a displacement of 4,369 tons (according to other sources, 3,420 tons), a length of 85 m, an underwater speed of up to 30 knots (~15 m/s) and having a maximum immersion depth of 400 m (see Fig. 11).

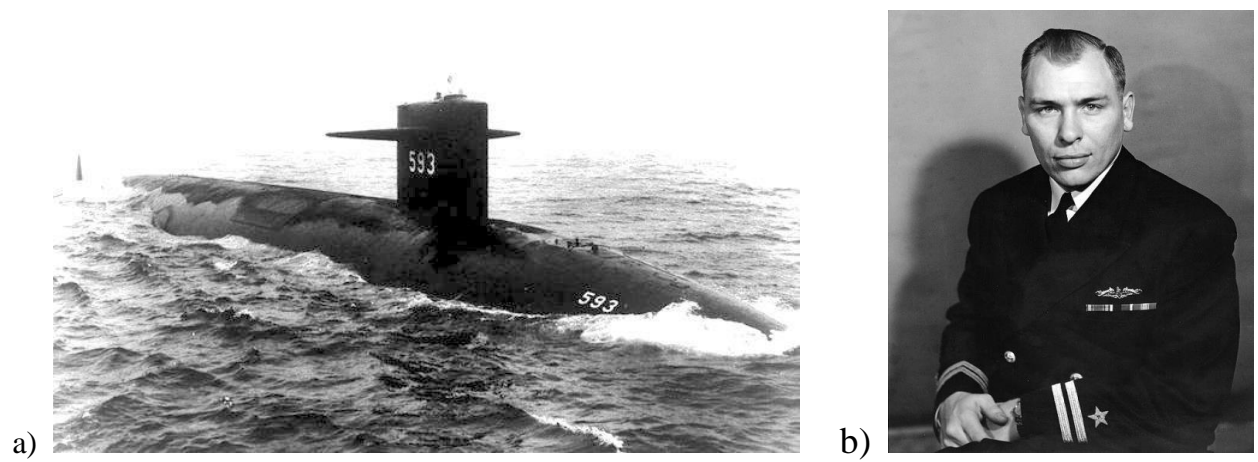

a) b)

*Fig. 11. a) Submarine "Thresher". b) A commander Lieutenant Commander John Wesley Harvey (1927–1963), graduate of the United States Naval Academy. Completed an additional training course on nuclear reactor control. He served on the nuclear submarine "Nautilus", the world's first nuclear submarine. Member of two Arctic expeditions of the submarine fleet to the North Pole. Appointed commander of the "Thresher" in January 1963. Photos are taken from the open sources on the Internet.*

By April 10, the submarine had passed all the tests scheduled after its repair with the exception of the last one – the deep diving. It should be noted that before the repair, "Thresher" had already sunk to a depth of 350 meters several times. This time there was another dive after repairs. Since it was a test trip, in addition to 16 officers and 96 enlisted, there were 17 civilian engineers and technicians from the Portsmouth docks on a board to observe the tests; the total of 129 people led by the Commander John Harvey. Further development of events took place in the following way.

On the morning of April 10, 1963, the "Thresher" sailed into the deep-water part of the Gulf of Maine accompanied by the rescue ship USS "Skylark" with which hydroacoustic communications were maintained. At 07:47 local time, "Thresher" began the dive; she had to spend about six hours under water. At 08:02 "Thresher" sent a message that the she had reached a depth of 120 meters; a solid hull, outboard fittings and pipelines were inspected. At 08.09 a message was received on the "Skylark" that half of the way to the planned dive depth was covered, the rate of diving was slowing down. At 8:35 the boat reached a depth 270 m that is 90 meters less than the limit. After another 18 minutes, the boat approached the maximum diving depth and a message was received from it that its course had not changed. At 09:09 the "Thresher" reached the maximum allowable diving depth of 360 meters. A minute later, a signal was sent to the submarine to change course, but the call was not answered. A minute later there was no answer and the repeated call.

At 9:13 am "Skylark" received a message: "*Experiencing minor problem. Have positive up angle.*" "*Attempting to blow. Will keep you informed.*" After a short time, noises were heard on the "Skylark" that were similar to those that accompany the release of compressed air. At 9:16 am, the escort ship "Skylark" received an indistinct message, from which it was possible to recognize only two words "test depth", and then at 9:17 am the last information arrived from the submarine, from which only two words could be made out: "900 ... north". Experts believe that "Thresher" provided location data. However, shortly thereafter, the acoustics on the "Skylark" heard a sound similar to the crackling sound of a ruptured, sturdy submarine hull. An alarm was announced in the fleet, a search and rescue operation were organized, but "Thresher" and its crew could not be saved ...

After a long search, the remains of the submarine, which had split into six large pieces within a radius of 300 meters, were recovered by means of the bathyscaphe "Trieste" at a depth of 2,560 meters southeast of Cape Cod, Massachusetts. According to experts, "Thresher" fell to the bottom with a stern down at a very high speed. On impact on the bottom, the submarine broke into pieces and plunged deep into the silt. In Fig. 12 one can see photographs of individual fragments of the sunken nuclear submarine. Many hypotheses were put forward about the reasons for the death of the submarine, among which there were suspicions that "Thresher" was somehow destroyed by a Soviet submarine (it must be borne in mind that these events took place in the midst of the Cold War). But let's approach the analysis of events and messages using established facts.

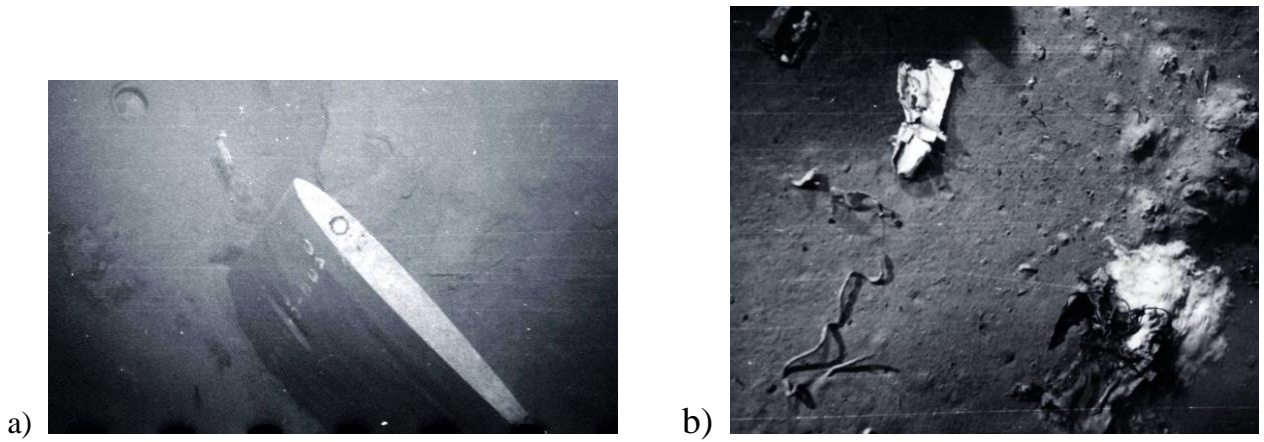

*Fig. 12. a) The top of the Thresher rudder with an element of navigation lights photographed by the bathyscaphe "Trieste" in October 1964. b) Elements of thermal insulation on a board of the "Thresher". Photos are taken from open sources on the Internet.*

In September 1963, American oceanographer Professor Columbus Iselin of the Woods Hole Oceanographic Institute, located near the Gulf of Maine, reported [19] that on April 8, two days before the disaster, a strong storm was observed in this area, which gradually shifted to the north-east and still did not subside by April 10, but was already over the Gulf of St. Lawrence (see Fig. 13). In the Gulf of Maine, the surface roughness had already considerably calmed down by this time, but the slowly moving internal waves induced by this storm could by the time of the sinking of "Thresher" be at the test site of the nuclear submarine at the exit from the Gulf of Maine. Based on the analysis of local hydrometeorological conditions, Iselin quite reasonably assumed that these internal waves could have large amplitudes, up to 100 m and more, and characteristic lengths of the order of 1–2 km. During the catastrophe, no hydrological observations were carried out; therefore, the given estimates are tentative, but very plausible.

According to Iselin's hypothesis, "Thresher" could have been lost due to a huge underwater vortex and the additional impact of an internal wave [19]. Red marks in Fig. 13 show the site of the submarine wreck, and the inset taken from Iselin's article [19] shows two underwater banks, Georges Bank and Browns Bank. Iselin assumed that the whirlpool could have been created by a past storm, which redirected the eastern current in the Browns Bank area to the south, whereas in normal weather it bypasses the bank in the northern part from east to west. This eastern current, meeting with the permanently presented current emanating from the Gulf of Maine between the Georges and Browns Banks, could have formed a vortex. The vortex motion of the liquid mass creates a decrease in the pycnocline, on which, possibly, an additional decrease was superimposed due to the 100-meter internal wave arriving at the same point. Further Iselin suggested that if, by coincidence, the "Thresher" found itself exactly in the place where the vortex and the internal wave that came up here were located, then it could be pulled into the depths due to a sharp drop in buoyancy. He also assumes that events were developing too rapidly, so that the crew did not have time to work out an emergency ascent.

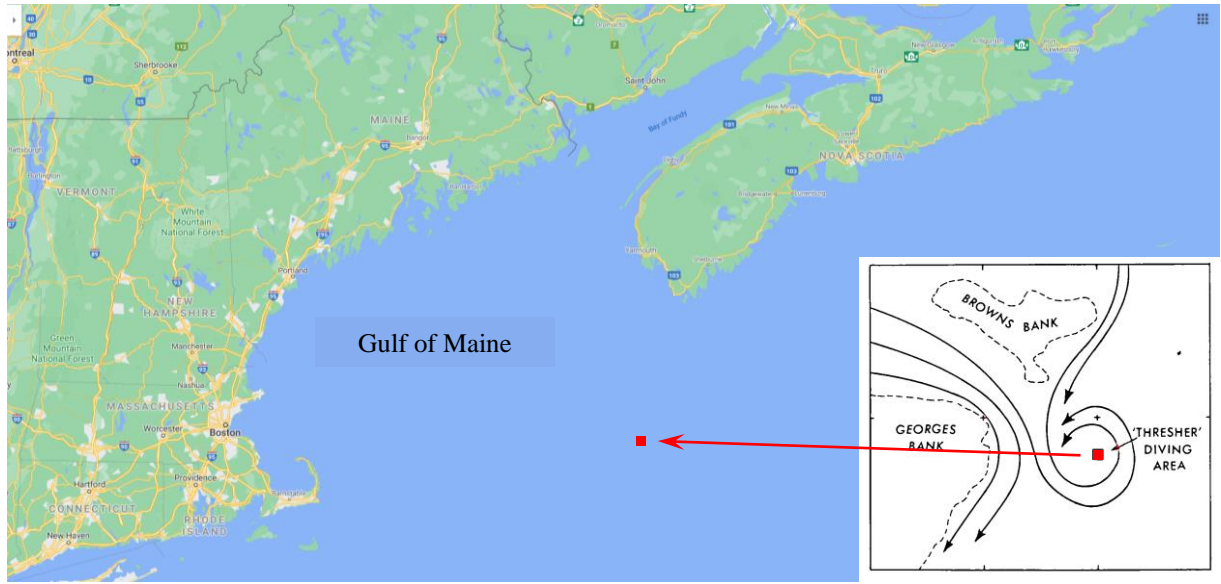

*Fig. 13. To the hypothesis of C. Iselin about the accident of the nuclear submarine "Thresher". The red squares mark the approximate place (41°46'N, 65°03' W) where the submarine was lost.*

Although Iselin's hypothesis looks physically consistent, it still contains too many assumptions. This applies to the unverified hypothesis of the formation of a whirlpool, to the assumption of a giant internal wave superimposed on the lowered pycnocline, to the assumption that the nuclear submarine was accidentally at an unnecessary time in the very epicentre of the danger zone. In addition, Iselin did not compare the reports received from the nuclear submarine with the alleged development of events. At the same time, he rightly noted that if "Thresher" experienced a failure in the operation of the ballast system, this could lead to tragedy.

In our opinion, the situation could be much simpler. Let's try to reconstruct the catastrophe on the basis of the available data without invoking the hypothesis of a giant whirlpool. At the time of the accident, "Thresher" had already reached the maximum allowable diving depth ~360 meters. This is the typical depth for the main pycnocline (see Fig. 3). Usually, during deep-sea diving, submarines move very slowly and carefully, the team listens to the creaks and crackles of a solid hull and fittings compressed by a pressure of ~36 atm (remember that the pressure in the water rises by one atmosphere after every 10 meters of depth). It can be assumed that the "Thresher" was moving at a minimum speed of about 2–4 knots (1–2 m/s) – this is a common practice during deep-sea test trials (Iselin assumed that the speed of the submarine was 5–6 knots [19], but, it is possible that "Thresher" did not move at all). Suppose that the submarine is overtaken by ISW in the form of a Gardner soliton with an amplitude of 100 m or more, moving at a speed of 2.5 m/s and having a shape of a trench as shown in the inset to Fig. 8a). This will cause the boat to sink to the stern and lose buoyancy, at first smoothly, and then more and more rapidly (see Fig. 14).

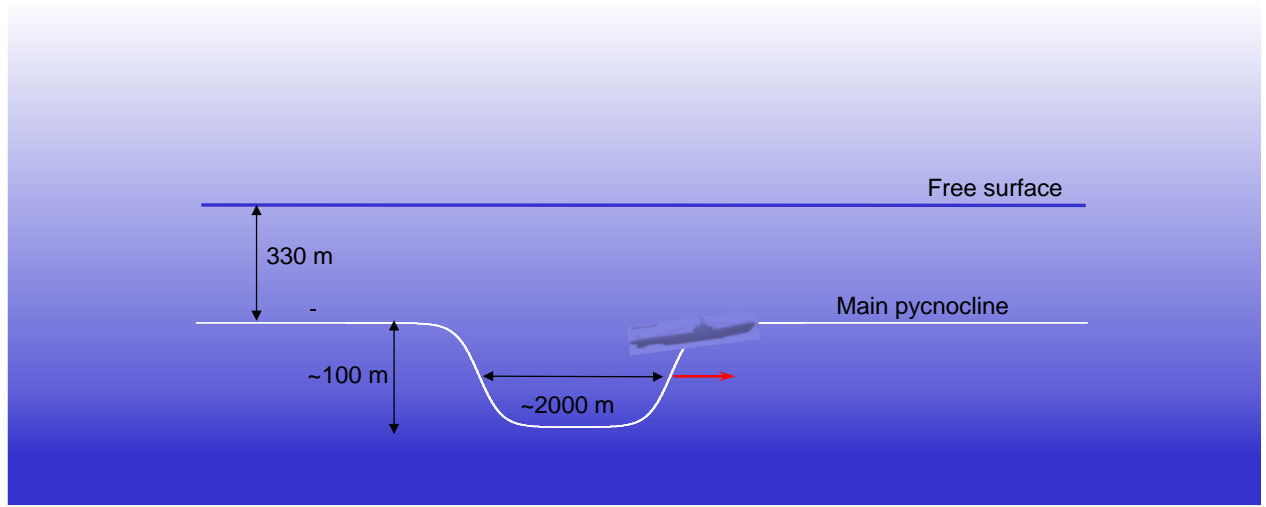

*Fig. 14. A hypothetical scheme of the accident that occurred with the nuclear submarine "Thresher". Reconstruction by the author.*

Let's listen to the words of the commander of "Thresher". The first thing he reports without much alarm is: "*Experiencing minor problem*". For the seasoned submarine commander, as John Harvey is characterized, this phrase sounds extremely unprofessional. The commander had to clearly report what the problem was, and not get off with a vague message about a "*small problem*". What could cause such an uncertainty?

Apparently, not at all because something happened onboard of the submarine, otherwise the commander would know exactly what to report. But if the submarine suddenly starts to sink at the stern for an incomprehensible reason for the commander, then he cannot report anything concrete, except to report on the very fact of the boat sinking ("*Have positive up angle.*"). It seems to the commander that the problem is not serious, which his crew will try to eliminate by blowing up the main ballast tanks, therefore he reports his actions: "*Attempting to blow. Will keep you informed*". All actions of the commander look completely correct and logical, given the lack of understanding of the essence of the problem (at that time, the existence of large-amplitude ISWs was not known). But the submarine is being pulled into the depths faster and faster. The blowing of the tanks does not keep pace, and the horizontal rudders are not effective at low speeds, the boat overshoots the maximum permissible depth, after which the collapse becomes inevitable. The crew had only a few minutes to eliminate the buoyancy deficit (see the above estimates of submarine immersion with a sharp change in water density) but, as it became known later during the investigation of this accident, the mechanism for purging the tanks during an emergency ascent was extremely imperfect; it did not allow the tanks to be blown out in a short time due to the icing of the valves caused by the rapid expansion of the highly compressed air. It turned out that the same problem was quite common in the American submarines of the previous generation.

Recently declassified documents about the sinking of the nuclear submarine "Thresher" confirmed a previously published finding presented in the official Report of the Commission of Inquiry into the Cause of the Submarine Crash that "*The **most likely** explanation [for the sinking of the submarine] is that the connection of the pipelines in the seawater system in the engine compartment broke through. The resulting spray shorted the electronics and triggered the automatic shutdown of the nuclear reactor*". However, in our opinion, this conclusion does not agree well with the reports of the submarine commander. For the sake of fairness, we note that the picky commission investigating the reasons for the death of the nuclear submarine, indeed, retroactively revealed a lot of violations, including very serious ones, during the repair of the boat at the shipyard in Portsmouth. One of the defects identified by the commission during the preparation of the boat for the cruise was the poor quality of welded work. However, according to the analysis of reports from the boat at the time of the accident, it seems that her death was due to another reason, not related to any damage on board. Indeed, first of all, the commander would hardly call a breakthrough of one of the pipelines with water entering one of the compartments, and even more so an accident with a nuclear reactor "*a small problem*". Secondly, he probably would have clearly reported in two words what happened. Thirdly, it is very doubtful that the failure of the reactor led to the subsidence of the boat. If one of the compartments began to fill with water due to a pipeline break, the commander would immediately report that the situation on board was near catastrophic, requiring an emergency ascent. And fourthly, when the reactor was muffled, the main ballast tanks could be blown out and surfaced in an emergency with the help of spare energy sources (electric batteries) that were on board the nuclear submarine. In the conclusions of the commission, it was noted that the power of the electric motor, backing up the reactor in emergency situations, was not enough in this case. Probably the engine had to be turned on in order to gain speed, which would make it possible to control the horizontal rudders of the boat, providing an ascent. At a certain angle of attack and speed of the boat, the lifting force acts on it, like an airplane. Logically, the capacity of the batteries had to be calculated to replace the failed reactor. But if the electric power was not enough to give the boat the required speed in a short time, then it obviously should have been enough to blow the main ballast tanks. To do this, it was necessary to release compressed gas from high-pressure cylinders, which, probably, could also be opened manually using valves. It was here that the commission revealed a serious flaw in the design of the ballast system, which did not allow for a quick purge of the main ballast tanks in order to prevent the boat from stalling in depth. Subsequently, the main ballast tank purging system was improved and replaced on all US nuclear submarines.

Probably, having picked up speed when falling, by inertia, the boat slipped through the level of neutral buoyancy in the centre of the ISW by tens or even hundreds of meters, where the external pressure was already prohibitive. Falling stern forward, the boat collapsed; its fragments were later found scattered within a radius of 300 m. On impact on the bottom, it fell apart. Naval historian Norman Friedman concluded from recently declassified commission materials that inadequate crew training exacerbated the problem, as the crew was unable to respond quickly enough to save the sub. However, the seamen cannot be blamed for the poor functioning of the ballast system. In those conditions, when the count of time went by seconds, the crew, presumably, did everything in their power, but the tragedy was not in their power to prevent. The ballast system was too sluggish. This is essentially confirmed by the author of an article in the American magazine Popular Mechanics Kyle Mizokami [20]: "*A trove of recently declassified files on the tragic 1963 sinking of the nuclear-powered attack submarine USS* Thresher *confirm the U.S. Navy didn't cover up the mysterious accident – and, in fact, there was no single event or error that caused the sub to sink*".

The possible role of internal waves in the destruction of the "Thresher" was discussed by many oceanographers (see, for example, [19]), but the authors limited themselves to only evaluative considerations about the physical possibility of the harmful influence of ISWs on the submarine, without in any way connecting the dynamics of the catastrophe development with the reports, coming from the boat in the last minutes before her death.

*The wreck of the nuclear submarine "Scorpion", May 1968*

The "Scorpion" was one of the US nuclear attack submarines with almost 8 years of service in the Navy (Fig. 15). The boat got its name in honour of the submarine of the same name "Scorpion", which mysteriously disappeared in 1944 during the Second World War. A similar fate befell this boat.

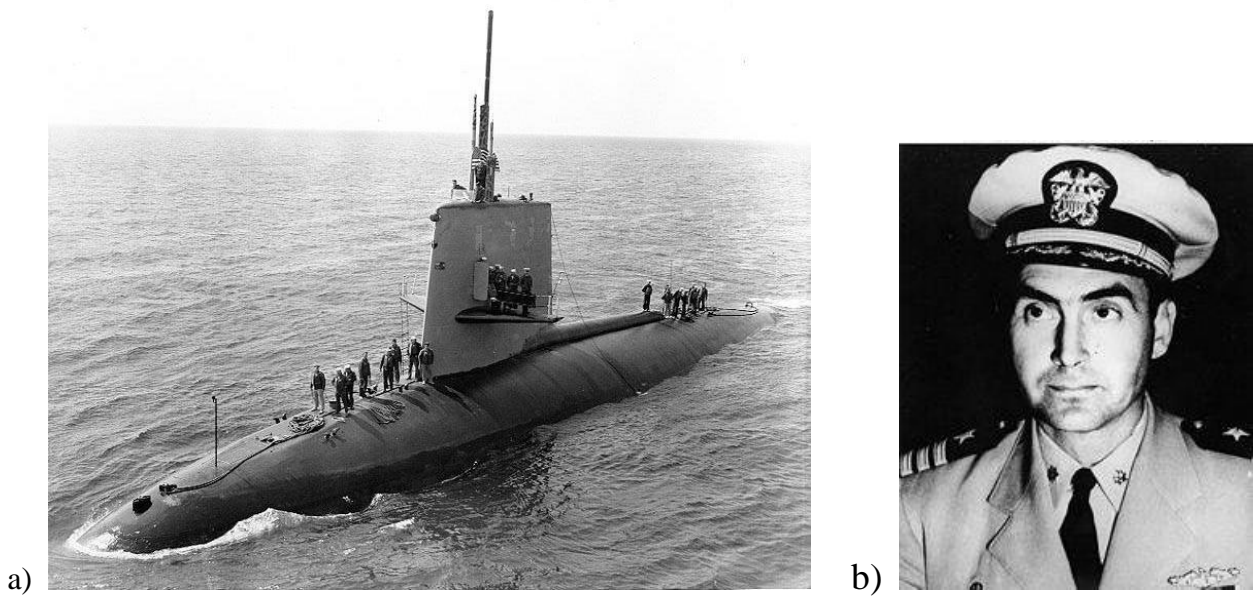

*Fig. 15. a) Nuclear submarine "Scorpion". b) Submarine commander Francis A. Slattery (1931–1968).*

In May 1968, the submarine was returning from combat duty in the Mediterranean to Norfolk base, USA. The last contact with the boat was on May 21; the commander reported that everything was in order on board, the boat was 80 km south of the Azores and was returning to base. But after the planned five days, the boat did not arrive at the base. An alert was announced for the fleet, the boat was searched by all available means; it was an unprecedented search and rescue operation. More than five months later, the site of the sinking of the nuclear submarine "Scorpion" was discovered by the triangulation method using the processing of acoustic signals from underwater hydrophones. Her remains rested at a depth of 3,047 m in the Mid Atlantic (see Fig. 16) and were later examined using the bathyscaphe "Trieste-2". All 99 crew members on board the nuclear submarine perished.

An authoritative commission was created to investigate the causes of the tragedy; the commission completed its work by the end of 1968 and stated that the submarine exceeded the maximum diving depth and sank "***for an unknown reason***". Dozens of versions of the death were considered, among which there were assumptions about a collision with a Soviet submarine or even an attack by a Soviet submarine. In the end, the commission came to the unexpected conclusion that the most likely reason was the death of the "Scorpion" from being hit by its own torpedo. According to this version, one of the torpedoes on board fell into disrepair (a similar event led to the death of the Russian nuclear submarine "Kursk" in August 2000). The commander ordered to shoot her overboard, but the homing torpedo, not finding a target, presumably turned around and aimed at its own nuclear submarine.

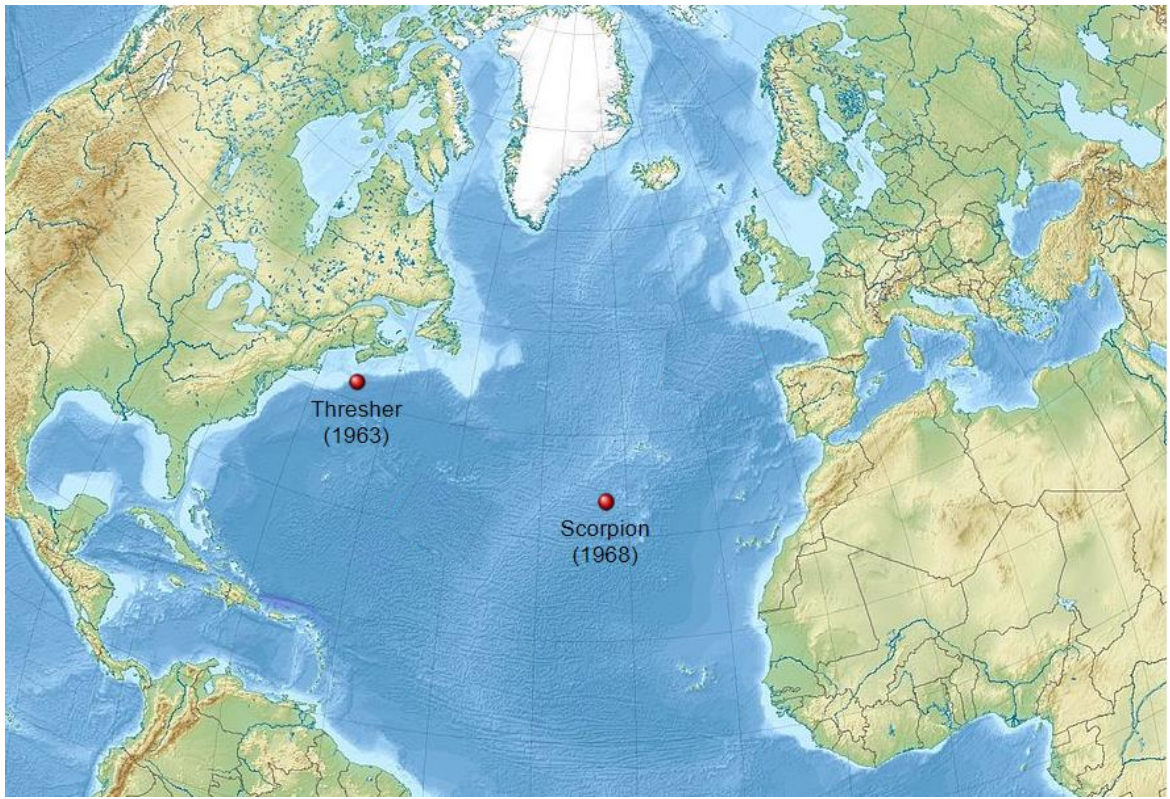

*Fig. 16. The places where the submarines "Thresher" and "Scorpion" were lost.*

"In 1970, a different Navy panel completed another classified report that disavowed the Court of Inquiry's conclusion. Instead of an accidental torpedo strike, the new group suggested a mechanical failure caused an irreparable leak that flooded the submarine." – https://www.chicagotribune.com/news/ct-xpm-1998-05-21-9805210149-story.html

Spokesman Commander Frank Thorp said after the investigation that: "While the precise cause of the loss remains undetermined, there is no information to support the theory that the submarine's loss resulted from hostile action or any involvement by a Soviet ship or submarine". The conspiracy theory that the Scorpion was attacked by a Soviet submarine recently surfaced again in a book "Scorpion Down" by a military columnist Ed Offley but none of the experts seem to take it seriously. One of the participants in the investigation, Ross Saxon, who examined the sunken boat from the bathyscaphe "Trieste-2" said that the boat did not fall apart into small pieces, but, apparently, its hull collapsed on impact on the bottom. *No traces of torpedo damage were visible on the hull* (see Fig. 17).

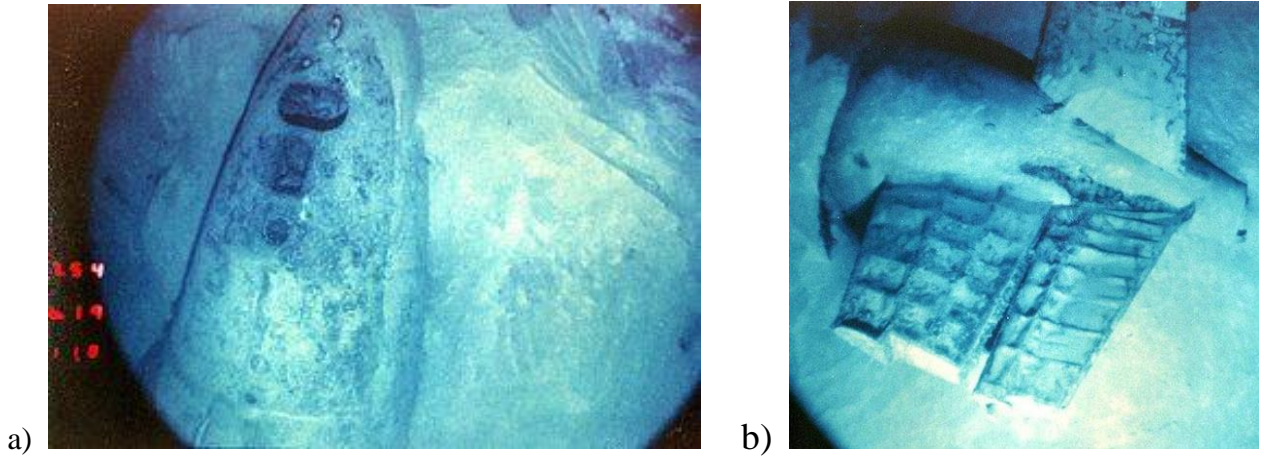

*Fig. 17. Bow (a) and aft (b) parts of the sunken nuclear submarine "Scorpion".*

During the investigation, a mysterious fact was revealed that there was the alternation of a series of about 17 underwater claps, following after some delay of the first clap on the recordings of acoustic sensors. These claps were recorded by the US sonar system in the area of the sinking of the boat. Logically, if it was associated with the collapse of the nuclear submarine, the source of the claps would have to approach the coast of the United States, but in fact, it was moving away. According to a study by the famous American scientist, a specialist in detecting objects lost in the ocean, John P. Craven, the impression was created that the boat turned 180° and moved away from the base. The explanation of this fact was associated with the assumption that the nuclear submarine tried to dodge the torpedo that was pursuing it and therefore turned in the opposite direction. But here one poorly substantiated assumption about a torpedo out of order is superimposed on another, associated with an unexpected manoeuvre of the nuclear submarine. Therefore, this whole hypothesis seems too unlikely in the opinion of many experts, especially since, as mentioned above, Ross Saxon, when examining the remains of the boat, *did not find any traces of damage from a torpedo*. Other speculative hypotheses of the death of the submarine, listed in [21], also turned out to be extremely unconvincing, therefore they are not presented here. Upon completion of the investigation on the loss of the nuclear submarine, it was concluded that the submarine, for *some unknown reason*, exceeded the maximum immersion depth, which led to its death. The US Navy command admitted that the death of the "Scorpion" had become a real mystery, which the commission of inquiry was never able to fully solve.

But let us imagine that the "Scorpion" overtook a large-amplitude ISW moving with the speed of 2.5 – 3.0 m/s. Such large ISWs were repeatedly recorded in the central Atlantic, including near the path along which the "Scorpion" was moving (see Fig. 16). In particular, in the work of the author with colleagues [22], an ISW was reported with the amplitude of 85 m in the region of the Guiana basin to the north-east of the Eastern Coast of South America. The maximum speed of the "Scorpion" was 30 knots (about 15 m/s), but the boat could move with much less speed, given its age and recent hastily completed repairs [21]. According to the technical characteristics, the boat's operating depth was 210 m, and the limiting depth was 300 m, but it is doubtful that the boat was sailing at the operating depth, because, as noted in [21], before the repair the boat was in such a deplorable state that its ballast system did not allow diving to more than 90 m. It is possible that after a hastily completed repair, the boat could dive to somewhat greater depths, but it is doubtful that it could withstand diving to such great depths as was planned during its construction. It is more likely that the boat was sailing at depth of the seasonal pycnocline ~(90 – 100) m.

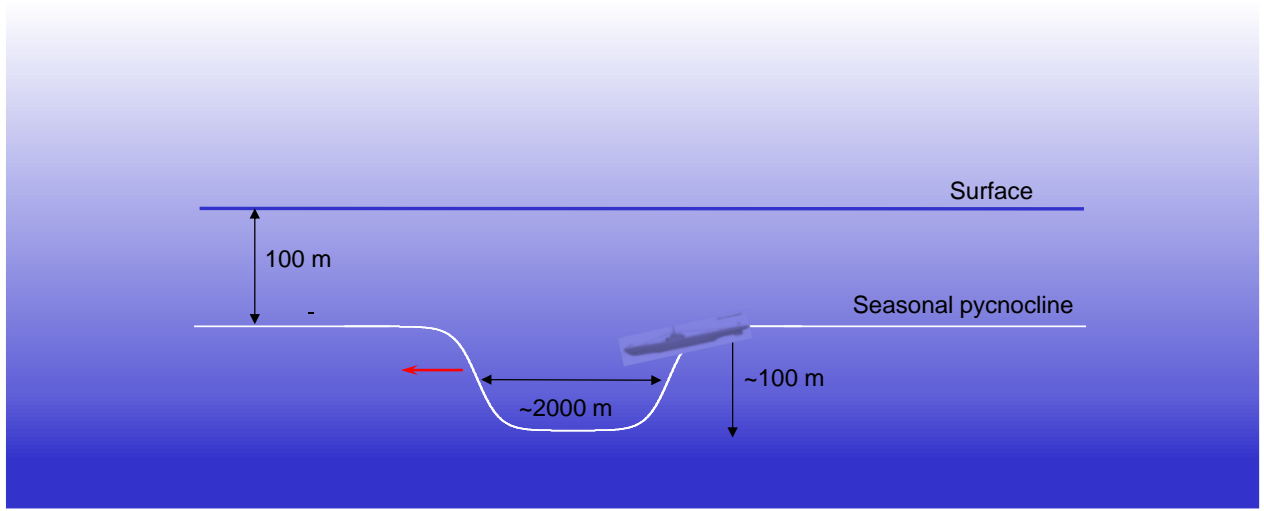

*Fig. 18. The hypothetical scheme of the accident that could occur with the nuclear submarine "Scorpion". Reconstruction by the author.*

At the indicated speed difference, catching up with the ISW and entering the area of reduced density, the submarine began to lose buoyancy, and not evenly, but as if sliding down a steep hill with its nose down (see Fig. 18). At the same time, the running engine also pushed the boat from behind. On a relatively steep slope of the ISW, the submarine could even flip or turn 180° from the intended course (imagine a log slipping down of a steep hill). The team had no time to level the course after such a somersault; it was necessary to urgently take measures for an emergency ascent, when every second is precious, so the boat continued to move in the opposite direction from the intended course. The time for an emergency ascent was sorely lacking, because according to available information, the ballast system on the boat was also outdated, like on the submarine "Thresher". As it was later found out, during the recent overhaul "on the "Scorpion", no alterations were made to the emergency blowing system of the main ballast tanks" [21]. As a result of the immersion, one of the compartments (probably the bow one) was the first to crack, and the sea water under high pressure $\sim(10-20)$ atm began to fill it quickly. Apparently, this first clap was later discovered in hydrophone recordings. But the boat was still alive, although it was already doomed, and continued to move in the opposite direction, fighting for survivability. The first compartment was filled with water and began to pull the boat further down. Then, one after another, the bulkheads between the compartments began to collapse with a crash, which led to a series of subsequent claps recorded by hydrophones. Of course, this is only the author's reconstruction of the reasons for the death of the nuclear submarine "Scorpion", but it is no more fantastic than all other versions, and, perhaps, even more realistic.

*The death of the submarine of the Indonesian Navy "Nanggala-402", April 2021*

Diesel submarine of the Indonesian Navy "Nanggala-402" was built in Germany in 1978 (Fig. 19). A boat with a crew of 53, led by Lieutenant Colonel Heri Oktavian, was on a training exercise in the Bali Sea. On the morning of April 21, 2021, the Indonesian naval forces lost contact with the boat shortly after it was given permission to dive deep for a torpedo firing exercise. Soon from one of the helicopters, and then from the search ships, a message was received about an oil slick in the alleged location of the submarine. Ships from different countries joined an intensive search for the missing submarine and after a few days her remains were found in the Bali Sea at a depth of 838 m. The boat split into at least three parts – the hull, the main part and the stern. A remote-controlled camera captured the final resting place of the submarine along with the crew on the seabed 96 km north of the Bali Island.

To the southeast of the place where the submarine perished, there is a strait about 40 km wide between the Bali Island and Lombok Island (see Fig. 20). In this place, intense currents and internal waves are regularly observed, which noticeably intensify every two weeks; one example is shown in Fig. 20. According to Bernadette Sloyan of the Australian research organization CSIRO, Tasmania Oceans and Atmosphere Division, "The Indonesian seas are a very small part of the ocean, but they account for about 10% of the tidal energy of the oceans" [22].

Representatives of the Indonesian Navy believe that the most likely reason for the death of a submarine is internal waves of large amplitude. This is the first time that the official authorities have confessed that internal waves could be a culprit in the submarine disaster. For the sake of fairness, we note that other possible reasons for the death of the submarine were also named (fire on a board, technical malfunction, crew errors), but they are still considered less likely.

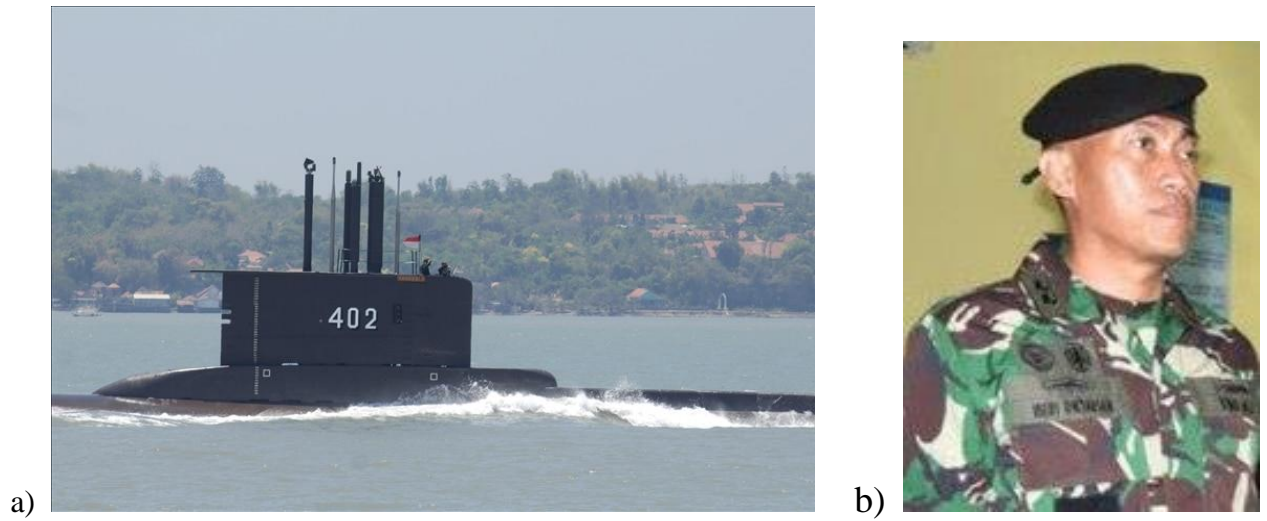

*Fig. 19. a) Indonesian submarine "Nanggala-402". b) The submarine commander Lt. Col. Heri Oktavian (1979–2021).*

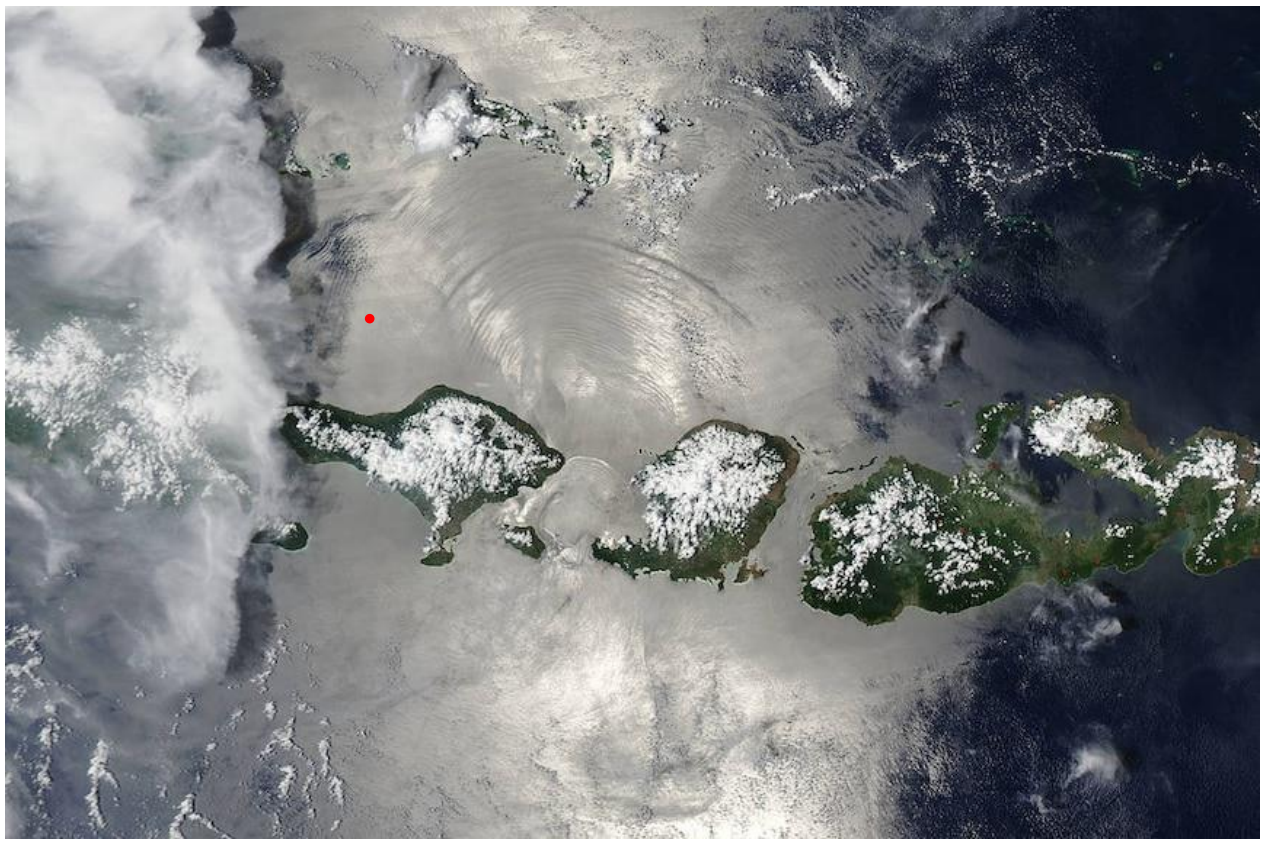

*Fig. 20. Satellite image of stripes formed on the surface of the Bali Sea by internal waves. Waves emanate from the Lombok Strait between Bali (left) and Lombok (right). The picture was taken in 2016 (NASA: Jeff Schmaltz, MODIS Land Rapid Response Team, NASA GSFC) [1]; such trains of internal waves regularly appear in this area. The red dot marks the approximate place of the sinking of the "Nanggala" submarine.*

The submarine parameters which are of interest to us are the underwater displacement of the boat was 1,390 tons, the working diving depth was 200 m, safe diving depth – 257 m, maximum diving depth – 500 m, the underwater speed was 25 knots (~12.5 m/s). A disaster with a submarine could develop according to the same scenario as in the case of the death of the nuclear submarine "Scorpion". If the boat were sailing at a great depth, as it was assumed under the conditions of the exercise, and encountered a high-amplitude ISW, then it could, as described above, "peck its nose" and fall 100 m lower, experiencing an additional load of almost 7 tons. The engines only increased the sinking effect, pushing the boat deeper (see Fig. 18), and the delay in rudder control and the inert ballast system made it impossible to cope with the problem.

**Conclusions**

Service on the submarine, even in peacetime, is still one of the riskiest activities. Despite the safety and comfort measures taken on board of modern large submarines, they are exposed to the dangers associated with insufficiently well-studied wave processes occurring in the ocean. Here they can be trapped by intense currents, vortices and large-scale internal waves, among which there may be so-called "rogue waves" that suddenly appear in certain places and just as suddenly disappear [24]. This type of waves on the surface of the ocean, which has destroyed and badly battered many surface ships, is now well known to oceanographers, although the theory is still far from complete. One of the most topical challenges is the problem of predicting the appearance of such waves and their parameters; many scientists in different countries are working on its solution. Much less is known about internal rogue waves. Unfortunately, in recent years, the activity of expeditionary work to study ocean waves and eddies in all countries has noticeably decreased. In the 70s – 80s of the last century, such work was intensively carried out in the USSR, USA and many other countries, which brought many unexpected discoveries and enriched the science of the ocean.

To reduce the risk of submarine navigation, it is necessary to continue cooperative research of scientists from different countries, allocate funding for expedition work, conduct ocean zoning in order to identify areas of especially intense internal waves, describe the typical parameters of internal waves in these areas, and collect statistical data. One of the first attempts to determine the statistical characteristics of ISWs was carried out in the work of the author and colleagues [16]. At the present time, atlases [6–8] contain rather extensive statistical material on ISWs in the World Ocean.

It is also necessary to develop means of detecting dangerous ISWs from the submarine or from aircraft and satellites in order to exclude a meeting with them. It is a challenge to scientists to develop of underwater remote sensing instruments for the spatially resolution of internal wave fields, so as to match the satellite imagery for the early detection of large ISWs. Probably, certain recommendations should be provided to submariners during the campaigns not to descend to great depths unless absolutely necessary, especially in peacetime (not discounting that such instruction perhaps already exist). From all of the above, it follows that internal waves pose a serious danger to submarine navigation, and the ocean does not forgive mistakes and neglect.

In conclusion, a brief summary of the largest losses of submarines of the USSR/Russia and USA is presented in Fig. 21. We only can wish to the submarines that remain "on Eternal Patrol" in the ocean, may their crews and civilian specialists Rest in Peace.

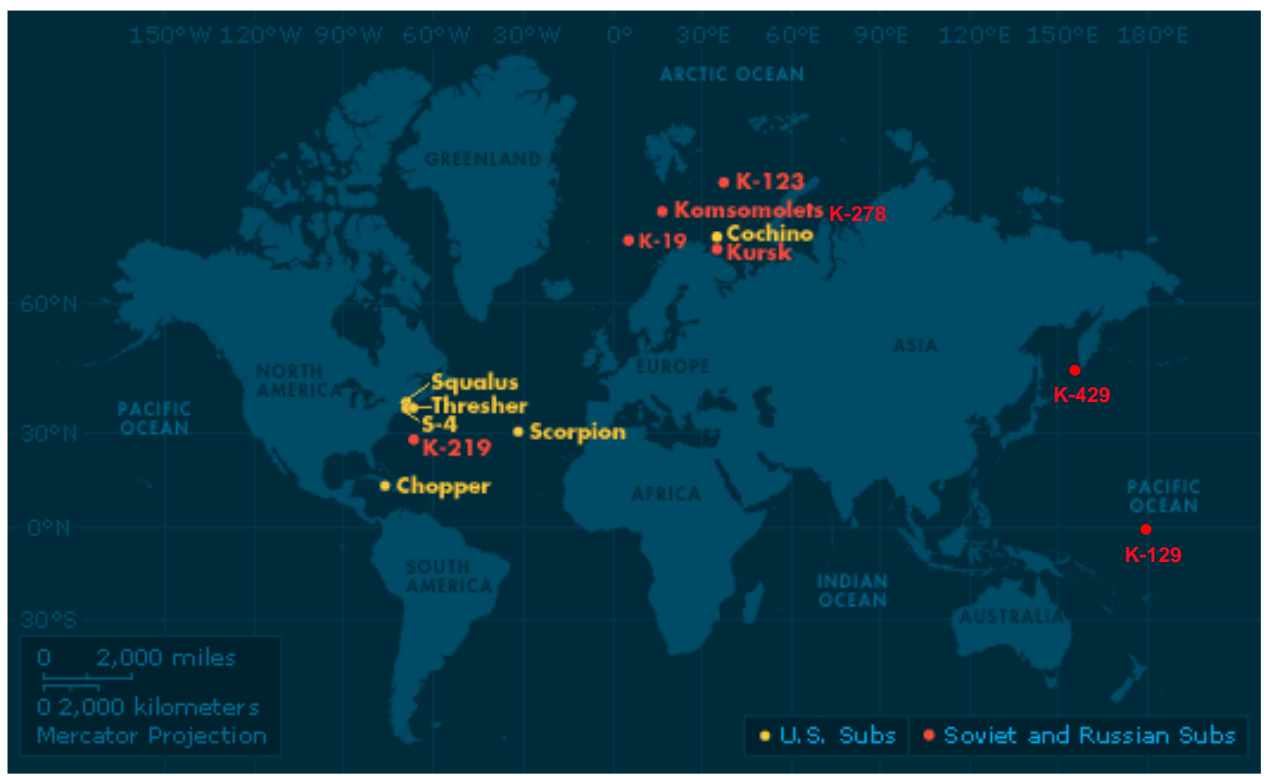

*Fig. 21. Dismal statistics of the largest losses of submarines of the USSR/Russia and the United States.*

**Acknowledgements.** The author is thankful to R. Addie, S. Anderson, L. Maas, L. Ostrovsky, E. Pelinovsky and T. Talipova for their comments and useful discussions. The author acknowledges the funding of this study provided by the State task program in the sphere of scientific activity of the Ministry of Science and Higher Education of the Russian Federation (project No. FSWE-2020-0007) and the grant of President of the Russian Federation for the state support of Leading Scientific Schools of the Russian Federation (grant No. NSH-2485.2020.5).